\documentclass[12pt]{article}
\usepackage{pdproc,epsfig}

\begin{document}

\def\gtwid{\mathrel{\raise.3ex\hbox{$>$\kern-.75em\lower1ex\hbox{$\sim$}}}}
\def\ltwid{\mathrel{\raise.3ex\hbox{$<$\kern-.75em\lower1ex\hbox{$\sim$}}}}

\renewcommand{\thefootnote}{\alph{footnote}}
  
\title{
A NEW FIT TO SOLAR NEUTRINOS\\
 USING EXTRA DIMENSIONS}

\author{DAVID O. CALDWELL}

\address{ Physics Department, University of California
  \\
 Santa Barbara, CA 93106-9530, USA\\
 {\rm E-mail: caldwell@slac.stanford.edu}}

%
%

\abstract{A neutrino mass-mixing scheme which explains qualitatively
all present evidence for neutrino mass (the solar and atmospheric
neutrino deficits, LSND, and hot dark matter), and also makes possible
heavy-element nucleosynthesis by supernovae, requires at least one
light sterile neutrino.  String-inspired models with sub-millimeter
extra dimensions provide naturally light sterile neutrinos, as is
needed to explain the solar $\nu_e$ deficit.  This bulk sterile
neutrino provides a better fit to the solar data than conventional
models by having vacuum oscillations of the $\nu_e$ to its zero mode
and MSW oscillations to its first few Kaluza-Klein modes.  While the
prediction of the Super-Kamiokande energy spectrum gives a fit
probability of 73\%, the superior energy resolution of SNO's
charged-current spectrum will determine whether this neutrino scheme
is correct and can demonstrate that an extra dimension of $\sim60\mu$m
exists.  Should this be the case, there are important implications
for supernovae, ultra-high-energy cosmic rays, double beta decay,
and dark matter.}
   
\normalsize\baselineskip=15pt

\section{Introduction}

The four-neutrino scheme in which the solar $\nu_e$ deficit is
explained by $\nu_e\rightarrow\nu_s$ (where $\nu_s$ is a sterile
neutrino), the atmospheric $\nu_\mu/\nu_e$ anomaly is attributed to
$\nu_\mu\rightarrow\nu_\tau$, and the heavier $\nu_\mu$ and $\nu_\tau$
share the role of hot dark matter was originally proposed\cite{fournu}
in order to explain those three phenomena.  Later the LSND
experiment,\cite{LSND} which observed
$\bar\nu_\mu\rightarrow\bar\nu_e$, provided a measure of the mass
difference between the nearly degenerate $\nu_e-\nu_s$ and
$\nu_\mu-\nu_\tau$ pairs and required the three mass differences that
were already present in that neutrino scheme.  Exactly this same
pattern of neutrino masses and mixings appears
necessary to allow production of heavy elements (A
$\buildrel >\over \sim$ 100) by type II supernovae.\cite{rprocess}
The neutrino properties required to ensure heavy-element
nucleosynthesis in the neutrino-heated ejecta of supernovae provide
independent evidence for (1) at least one light sterile neutrino,
(2) a near maximally-mixed $\nu_\mu$--$\nu_\tau$ doublet split from
the lower mass $\nu_e$--$\nu_s$ doublet, (3) $\nu_\mu$--$\nu_e$
mixing $\gtwid10^{-4}$, and (4) a splitting between the doublets
(measured by the $\nu_\mu$--$\nu_e$ mass difference) $\gtwid1$
eV$^2$, favoring the upper part of the LSND range.

While qualitatively this neutrino scheme seems to explain all existing
neutrino phenomena, solar neutrino observations are now sufficiently
constraining that even the most viable $\nu_e\to\nu_s$ explanation,
a small-angle MSW transition, appears to be in difficulty.\cite{ref:4}
Although providing better fits to the solar data, even active-active
transitions in a three-neutrino scheme do not give a quantitatively
good explanation of those data.  In this paper we point out that
there is a way to achieve an excellent fit and rescue the apparently
needed four-neutrino scheme if large extra dimensions exist.   This
is motivated by the latest developments in string theories, which
have made plausible the interesting possibility\cite{nima} that there are
such large extra space dimensions
(\lower1ex\hbox{$\buildrel < \over \sim\/$}mm).

To strengthen the motivation for introducing such an exotic solution to the
solar neutrino problem, the next section deals with the particle physics
evidence for the four-neutrino scheme, with particular emphasis on the new
and much more compelling analysis of the LSND experiment,\cite{LSNDnew} and
the section
after that treats the need for the same neutrino scheme to make possible
the supernova r process.  Then two extra-dimension models are briefly and
qualitatively introduced, one having the usual low fundamental string scale
of a few TeV, and the other having a much higher scale ($\sim10^6$ TeV),
which has desirable phenomenological, and probably theoretical, advantages.
Next, the solar data are fit using a scheme which either model justifies.
While active neutrinos are confined to a brane, the sterile neutrino
occupies the bulk, the large size of which is responsible for the suppression
of its mass, solving an inherent problem of sterile neutrinos.  The small
mass difference between the $\nu_e$ and the zero mode of the sterile
neutrino causes vacuum oscillations, while low Kaluza-Klein excitations,
which have a mass of the inverse extra dimension size, give small-angle
MSW oscillations.  The parameters of the model are determined by the
average rates of the three types of solar experiments, and then the
energy spectrum is predicted.  This prediction agrees well (73\% probability)
with the Super-Kamiokande spectrum,\cite{ref:4} but the crucial test will
come from SNO, which will have far better energy resolution.  The model
parameters, which give quite different mass-difference/mixing-angle
regions from usual vacuum or MSW fits, provide important implications for
gravity experiments, double beta decay, ultra-high-energy cosmic rays,
supernovae, and dark matter, and these issues occupy the last section.

\section{LSND's Added Support for the Four-Neutrino Scheme}
Although not the original motivation for the four-neutrino
scheme,\cite{fournu} the three very different mass differences required
by the solar and atmospheric anomalies and the LSND experiment\cite{LSND}
is usually cited as the reason to have, in addition to the three light
active neutrinos allowed by the $Z^0$ width, a light neutrino ($\nu_s$)
not having the usual weak interaction.  As evidence for neutrino oscillation
explanations for the solar and atmospheric phenomena have become stronger,
it is essential to evaluate the LSND result.  A new analysis\cite{LSNDnew}
of data from
that experiment strengthens the conclusion that $\bar\nu_\mu\to\bar\nu_e$
oscillations were seen via events of the type $\bar\nu_ep\to e^+n$, with
the $e^+$ being identified by both scintillation and Cherenkov light from
a 167-metric-ton tank of doped mineral oil, and the $n$ detected by the
2.2 MeV $\gamma$ from $np\to d\gamma$.

The 1996 publication\cite{LSND} by LSND claiming a positive result was
based on data collected in 1993--5 using a water target in which 798-MeV
protons produced mainly $\pi^+$, of which 97\% were brought to rest,
providing a $\bar\nu_\mu$ beam from the subsequent decay at rest of the
$\mu^+$.  During 1996--8, data were obtained at a lower rate in parasitic
operation using a high-Z target.  The latter data sets had larger
cosmic-ray backgrounds, and conference reports using them had some
disturbing distributions, probably indicating a problem of accidental
electron-$\gamma$ coincidences.  This was likely because, along with
the higher background, the $R$ distribution (a measure of the likelihood
that the $e$--$\gamma$ was real as opposed to accidental) was the one
published distribution---not any event spatial distributions, as often
supposed---which was statistically worrisome.

The new analysis deals with all the data, and the various distributions
from the 1996--8 period all agree well with those from 1993--5, since
now the $R$ distribution gives a smooth fit to the data over the whole
energy range from 60 even down to 20 MeV.  From 20--36 MeV there is
added background from accidentals from $\nu_e\/^{12}{\rm C}\to e^-X$,
but this now gives no problem.  The $R$ value is obtained using laser
events for accidentals and Monte Carlo data for reals, checked by
cosmic-ray neutron events.  Three ingredients go into $R$: the $\gamma$
energy, the distance between the $e$ and the $\gamma$, and the neutron
capture time.

In the new LSND analysis\cite{LSNDnew} a simultaneous likelihood fit of
an event is made
to the position, energy, track direction, track length, and fraction of
Cherenkov (vs.\ scintillation) light utilizing the light amplitudes and
arrival times at each of the 1220 8{\tt "} photomultipliers looking into
the tank of mineral oil and scintillator.  The light amplitude and times
were formerly used separately in determining these output quantities.
A systematic skewing of event positions and angles was eliminated by
taking into account an exponential tail on the Gaussian charge
distribution from the photomultipliers.  The improvement in position
resolution in the new analysis reduced the most likely $e$--$\gamma$
distance from 74 to 55 cm, and the accidental $\gamma$ rate is proportional
to the cube of this distance.  The 1996 publication\cite{LSND} used a
cut at $R>30$, whereas the new analysis\cite{LSNDnew} uses $R>10$,
where for a given
analysis the larger the value of $R$ the more likely the $e$ and $\gamma$
are correlated.  For those two $R$ cuts from different analyses the
accidental rate has
decreased from 0.6\% to 0.3\%, while the correlated efficiency has
increased from 23\% to 39\%.  Using the $R>10$ cut there were 86 beam-on
events, $36.9\pm1.5$ beam-off background (which has a small error, since
data are collected during 94\% of the time between pulses), $16.9\pm2.3$
expected $\nu$ background, giving an excess of $32.2\pm9.4\pm2.3$ events.
The probability that $36.9+16.9=53.8$ background events fluctuate up to
the observed 86 is
$<1\times10^{-4}$, taking into account all errors.  A different way to state
the result is to fit the $R$ distribution, instead of using a cut.  This
gives $87.9\pm22.4\pm6.0$ events above expected backgrounds and corresponds
to an oscillation probability of $(0.264\pm0.067\pm0.045)$\%, which is
consistent with, but smaller than, the 1996 result\cite{LSND} of
$(0.31\pm0.12\pm0.05)$\%.

The results on $\bar\nu_\mu\to\bar\nu_e$ are quite compelling, and confidence
in them is increased by the excellent agreement with results on
conventional processes, such as $\nu_e$C and $\nu_\mu$C to ground and
excited states.  Unfortunately it was decided to make the analysis ``global''
and use not only the 20--60 MeV data for decay-at-rest
$\bar\nu_\mu\to\bar\nu_e$, but also the 60--200 MeV data for $\pi^+$
decay-in-flight $\nu_\mu\to\nu_e$.  The latter process has only one signal,
since it is detected by $\nu_e\/^{12}{\rm C}\to e^-X$, and hence suffers
from much higher backgrounds.  Furthermore, unlike the
$\bar\nu_\mu\to\bar\nu_e$ case for which large numbers of electrons from
cosmic $\mu$ decays provide calibrations and optimization of the analysis,
the higher energy $\nu_\mu\to\nu_e$ has no reference process and must
depend upon uncertain extrapolations from lower energy.  Thus the
published\cite{ref:6} LSND analysis of $\nu_\mu\to\nu_e$ was extremely
complex, and while yielding confirmatory evidence for oscillations
($18.1\pm6.6\pm4.0$) events above background, or an oscillation probability
of ($0.26\pm0.10\pm0.05$)\%, the results could never be used on their own
to claim observation of oscillations.  As might be expected, the ``global''
analysis---which is optimized for $\bar\nu_\mu\to\bar\nu_e$---sees no
significant evidence for $\nu_\mu\to\nu_e$, yielding ($8.1\pm12.2\pm1.7$)
events above
background, or an oscillation probability of ($0.10\pm0.16\pm0.04$)\%.

The lack of sufficient $\nu_\mu\to\nu_e$ events skews the results when
expressed in a mass-squared-difference ($\Delta m^2$), mixing-angle
($\sin^22\theta$) plot, favoring lower $\Delta m^2$ values.  The often
neglected island around 5--7 eV$^2$ is diminished in probability, whereas
in the published\cite{ref:6} $\nu_\mu\to\nu_e$ case it was actually the
most favored region.  Probably a better plot to use, even if out of date,
is that of Eitel,\cite{Eitel} which analyzes together LSND and
KARMEN\cite{KARMEN} results.  This incredibly computer-intensive work
has been not entirely correctly utilized in the new LSND plot.  Eitel's
work still gives the most representative picture if shifted to slightly
smaller mixing angles, corresponding to the LSND oscillation probability
shift of 0.31 to 0.26\%, since KARMEN now has more events said to be
consistent with background.  Except for some gaps, such as around 3--5
eV$^2$, all $\Delta m^2$ values from 0.2 to 10 eV$^2$ are possible.

\section{Supernova Evidence for the Four-Neutrino Scheme}

There was an apparent conflict between the production of the heaviest elements
in the neutrino-heated material ejected relatively long ($\sim10$s) after the
explosion of a Type II or Type I b/c supernova and at least the larger
$\Delta m^2$ values from LSND.  Limits were placed by this r process of rapid
neutron capture on $\nu_\mu$--$\nu_e$ mixing because energetic $\nu_\mu$
($\langle E\rangle\approx25$ MeV) coming from deep in the supernova core
could convert via an MSW transition to $\nu_e$ inside the region of the
$r$-process, producing $\nu_e$ of much
higher energy than the thermal $\nu_e\ (\langle E\rangle\approx11$ MeV).  The
latter, because of their charged-current interactions, emerge from farther out
in the supernova where it is cooler.  Since the cross section for $\nu_en\to
e^-p$ rises as the square of the energy, these converted energetic $\nu_e$
would deplete neutrons, stopping the $r$-process. 
Calculations\cite{ref:9} of this effect limit $\sin^22\theta$ for
$\nu_\mu\to\nu_e$ to $\ltwid10^{-4}$ for $\Delta m^2_{e\mu}\gtwid2$ eV$^2$, in
conflict with at least the higher mass region of the LSND results.

More recently, serious problems have been found with the $r$ process itself.
First, simulations\cite{ref:10} have revealed the $r$-process
region to be insufficiently neutron-rich, since about $10^2$ neutrons are
required for each seed nucleus, such as iron.  This was bad enough, but the
recent realization of the full effect of $\alpha$-particle formation has
created a disaster for the $r$ process.\cite{ref:11}  At a radial region
inside where the $r$ process should occur, all available protons swallow
up neutrons to form the very stable $\alpha$ particles, following which
$\nu_en\to e^-p$ reactions reduce the neutrons further and create more protons
which make more $\alpha$ particles, and so on.  The depletion of neutrons
rapidly shuts off the $r$ process, and essentially no nuclei above $A=95$ are
produced.

To solve this problem the $\nu_e$ flux has to be removed before the
$r$ process site, while leaving a very large $\nu_e$ flux at a
smaller radius for material heating and ejection.
The apparent miracle of having a huge $\nu_e$ flux disappear before it
reaches the radius of the supernova where $\alpha$ particles form can be
accomplished\cite{rprocess} if there is (1) a sterile neutrino, (2)
approximately maximal $\nu_\mu\to\nu_\tau$ mixing, (3) $\nu_\mu\to\nu_e$
mixing $\gtwid10^{-4}$, and (4) an appreciable ($\gtwid1$ eV$^2$) mass-squared
difference
between $\nu_s$ and the $\nu_\mu$--$\nu_\tau$.  This is precisely the
neutrino mass pattern required to explain the solar and atmospheric anomalies
and the LSND result, plus providing some hot dark matter!

Such a mass-mixing pattern creates two level crossings.  The inner one,
which is outside the neutrinosphere (beyond which neutrinos can readily
escape) is near where the $\nu_{\mu,\tau}$ potential $\propto(n_{\nu_e}-n_n/2)$
goes to zero.  Here $n_{\nu_e}$ and $n_n$ are the numbers of $\nu_e$ and
neutrons, respectively.  The $\nu_{\mu,\tau}\to\nu_s$ transition which occurs
depletes the dangerous high-energy $\nu_{\mu,\tau}$ population.  Outside
of this level crossing, another occurs where the density is appropriate for a
matter-enhanced MSW transition corresponding to whatever $\Delta m^2_{e\mu}$
LSND is observing.  Because of the $\nu_{\mu,\tau}$ reduction at the first
level crossing, the dominant process in the MSW region reverses from the
deleterious $\nu_{\mu,\tau}\to\nu_e$, becoming $\nu_e\to\nu_{\mu,\tau}$ and
dropping the $\nu_e$ flux.  For an
appropriate value of $\Delta m^2_{e\mu}$, the two level crossings are
separate but sufficiently close so that the transitions are coherent.  Then
with adiabatic transitions (as calculations show) and maximal
$\nu_\mu$--$\nu_\tau$ mixing, the neutrino flux emerging from the second
level crossing is 1/4 $\nu_\mu$, 1/4 $\nu_\tau$, and 1/2 $\nu_s$, and no
$\nu_e$.

Note that the $\bar\nu_e$ flux is unaffected at the level crossings,
so $\bar\nu_ep\to e^+n$ enhances the neutron number in the $r$ process region,
since the protons have not been depleted by $\alpha$ particle formation.
It should be emphasized that this mechanism is quite robust,
not depending on details of the supernova dynamics, especially as it occurs
quite late in the explosive expansion.

It is essential that the two level crossings be in the correct order, and this
provides a requirement on $\Delta m^2_{e\mu}$, since the MSW transition
depends on density and hence on radial distance from the protoneutron star.
Detailed calculations have been made for $\Delta m^2_{e\mu}\sim6$ eV$^2$,
which works very well.  Possibly $\Delta m^2_{e\mu}$ as low as 2 eV$^2$ or
maybe even 1 eV$^2$ would work, but that is speculative.  At any rate, the
mass difference needed in this scheme, which is the only one surely consistent
with all manifestations of neutrino mass and which rescues the $r$
process,\cite{ref:12} implies appreciable hot dark matter.

\section{Invoking Extra Large Dimensions}
An idea which has caused a lot of interest lately is that one or more of the
extra dimensions required by string theories may be of a size which is
observable and which would have a lot of consequences for our present universe.
The possible observations, however, seem to require higher accelerator
energies than presently available or difficult gravity experiments at
senstitivities not yet reached.  Instead, the case is made here that existing
data, or that available soon, can show evidence for extra dimensions in an
unexpected sector, the observation of neutrinos from the sun.

While it is said qualitatively that $\nu_e\to\nu_s$ can explain the solar
$\nu_e$ deficit by either small-angle MSW or vacuum oscillations,
quantitatively the solar experiments are so numerous and precise that the fits
to either solution are rather poor.  Even three-neutrino schemes, which use
active-active transitions (e.g., $\nu_e\to\nu_\mu$), do only marginally better.
One can be easily misled by these fit results because a very bad fit to one
type of data (e.g., the rates of the three types of experiments) can get
ignored when its few degrees of freedom are insignificant when included in
the fit are the many degrees of freedom from an energy spectrum,
zenith-angle data, etc.

If there are large extra
dimensions, then $\nu_s$ becomes a particle which can exist in the
extra dimension(s), or in other words it inhabits the bulk, while active
neutrinos are confined to the brane.  There could be more than one extra
dimension and more than one brane, but for simplicity the discussion here is
limited to one of each.  A characteristic of a bulk particle (such as the
graviton) is that it is really a series of states; this Kaluza-Klein tower
has mass values $m_n\approx n/R$, where $R$ is the size of the extra
dimension.

While many papers have been written about bulk sterile neutrinos, since
this provides a means of getting sterile neutrinos of small mass which have
some mixing with active neutrinos, generally these theories do not produce
three $\Delta m^2$ or attempt to explain all evidence for neutrino mass.  The
two models described here briefly and just qualitatively were developed by
R.N.~Mohapatra, whose contribution to these Proceedings goes into more detail.
These models provide the desired phenomenology to have both vacuum and
MSW oscillations for solar $\nu_e\to\nu_s$, while also giving suitable
masses and mixings for the active neutrinos compatible with the four-neutrino
scheme motivated above.  One model has the usual string scale of a few
TeV,\cite{ref:13} necessitating at least two extra large dimensions (although
only one enters the considerations here), while the other has a very high
string scale, $\sim10^6$ TeV, and hence only one extra dimension need be large.

The low-scale model was developed first, and the fit to the solar
data\cite{ref:14} is described in this basis, so somewhat more information
on this is provided here.  It is based on a mechanism\cite{ref:15} in which
one or more gauge singlet neutrinos in the bulk couple to lepton doublets
in the brane, and after electroweak symmetry breaking this coupling leads
to Dirac neutrino masses which are suppressed by the ratio $M_*/M_{P\ell}$,
where $M_{P\ell}$ is the Planck mass and $M_*$ is the string scale.  This is
sufficient to explain small neutrino masses and owes its origin to the large
bulk volume that suppresses the effective Yukawa couplings of the Kaluza-Klein
modes of the bulk neutrino to the brane fields.  In this class of models,
naturalness of small neutrino mass requires that one must assume the existence
of a global B-L symmetry in the theory, since that will exclude the
undesirable higher dimensional operators from the theory.  In particular this
leads to a neutrino mass $m_{\nu} = h {M^*v\over M_{P\ell}}\sim 10^{-5}$ eV,
where $\nu$ is the scale of SU(2)$_L$ breaking, and $h$ is the Yukawa coupling.
The $\nu_e$ and $\nu_s$ (zero mode, or ground
state, hence really $\nu_{s,o}$) are two, two-component spinors that form
the Dirac fermion with mass $m_\nu$.

In order to fit neutrino data, one needs to include new physics in the
brane that will generate a Majorana mass matrix for the three standard
model neutrinos of the form $\delta_{ab}$ (where $a,b= e,\mu,\tau$).
We assume that $\delta_{\mu\tau}$ is much bigger than the
other elements. As a result the $\nu_{\mu,\tau}$ in effect decouple from
the $\nu_{e,s}$ and do not affect the
mixing between the bulk neutrino modes and the $\nu_e$. Further, it leads
to maximal mixing in the $\mu-\tau$ sector as is needed to understand the
atmospheric neutrino data. If we choose $\delta_{\mu\tau}\sim $eV, then
this provides an explanation of the LSND observations. In the rest of this
discussion we focus only on the $\nu_e-\nu_s$ sector and how we fit
the solar neutrino data.

The new physics chosen to accomplish this is to introduce two singlets to
the Higgs sector on the brane to produce Majorana neutrino masses by radiative
effects at the two-loop level.  The radiative effects also split the
$\nu_e$--$\nu_{s,o}$ Dirac neutrino into two Majorana fermions, introducing
a very small mass difference;
the two mass
eigenstates are maximally mixed and have the values $\nu_{1,2}\approx
(\nu_e\pm\nu_{s,o})/\sqrt2$.  Thus as the $\nu_e$ produced in a weak
interaction process evolves, it oscillates to the $\nu_{s,o}$ state with an
oscillation length of the order of the Sun-Earth distance, giving vacuum
oscillations (VO).  Since the $\nu_e$ mixes also with the Kaluza-Klein (KK)
modes of the bulk neutrinos with a $\Delta m^2\sim10^{-5}$ eV$^2$, this brings
in the MSW resonance transition of $\nu_e$ to $\nu_{s,n}$ modes at higher
energies.

In this model the mixing angle is given by $\cos\theta=1/N$, where
$$N^2=1+(\pi m_o R)^2+(m_n/m_o)^2$$
for the $n$th KK state.  Since $m_oR\ll1$, the second term can be
neglected.  For the zero mode, $N^2\approx1+(m_o/m_o)^2=2$,
or $\theta\approx\pi/4$, which is the maximal
mixing required by experiment.  For $n\ge1$, $m_n\approx n/R$, so $N^2\approx1
+(n/m_oR)^2\approx(n/m_oR)^2$, which gives small mixing angles.

A second way to achieve the same phenomenology is possible using a much
higher string scale.  In this class of models,\cite{ref:16} one postulates
that the theory in the brane is left-right symmetric so that it contains
B-L as a local symmetry, which is more desirable than having it as a global
symmetry.  The gauge group of the model is SU(2)$_L\times$SU(2)$_R\times$
U(1)$_{B-L}$ with quark and lepton doublets assigned as usual in a left-right
symmetric manner and Higgs fields belonging to a didoublet field
$\phi$(2,2,0) and B-L doublets $\chi_{L,R}$.  The use of this model to give
suitable active neutrino masses (now through a type of see-saw mechanism)
and mixings and to produce the same phenomenology as in the low-scale model
cannot be described in a few lines, but it is spelled out in Ref.~18.  The
point does need to be emphasized here, however, that if there is only one
large extra dimension, $R$, and all small extra dimensions are $\sim M^{-1}_*$,
then $M^2_{Pl}\approx M^3_*R$ leads to $M_*\sim10^6$ TeV if
$R\sim\rm mm$.\cite{ref:16}

\renewcommand{\thefootnote}{\fnsymbol{footnote}}
\section{Fitting the Solar Data\footnote{Converting the theory to a fit
was done by S.J.~Yellin.}}

\begin{figure}[t]
\epsfxsize=3in
\begin{center}
\leavevmode
\epsfbox{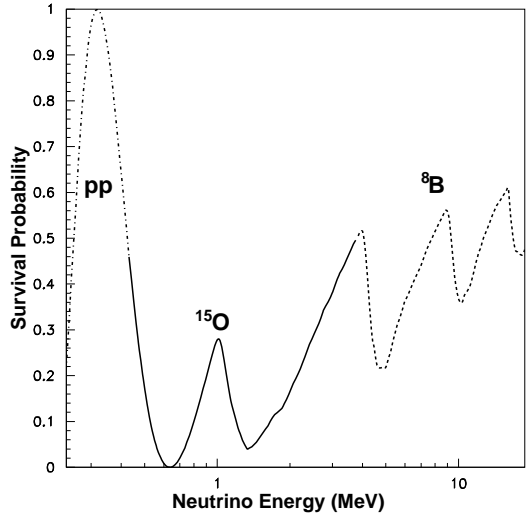}
\end{center}
\caption{Energy dependence of the $\nu_e$ survival probability when
$R=58\mu$m, $mR=0.0094$,
$\delta_{ee}=0.84\times 10^{-7}$ eV.  The dot-dashed part of the curve
assumes the radial dependence in
the Sun for neutrinos from the pp reaction, the solid part assumes $^{15}$O
radial dependence, and the dashed part assumes $^8$B radial dependence.}
\label{fig:edep}
\end{figure}

The models provide a naturally small mass difference between the mass
eigenstates formed from the $\nu_e$ and the $\nu_{s,0}$, giving vacuum
oscillations.  The first node of the survival probability function due to
VO can be used to suppress $^7$Be neutrinos.  Going up in energy toward
$^8$B neutrinos, the survival probability, which in the VO case would have
risen to very near one, is suppressed by the
small-angle MSW transitions to the different KK excitations of the bulk
sterile neutrinos, as is clear from Fig.~\ref{fig:edep}.
This is a new way to fit the solar neutrino data in models with
large extra dimensions and is the main observation of this report.

For comparison with experimental results, tables of detector sensitivity
for the Chlorine and Gallium experiments were taken from Bahcall's web
site.\cite{BP98}  The Super-Kamiokande detector sensitivity\cite{Nakahata}
was used with appropriate smearing of the neutrino-electron elastic scattering
cross section.  Calculations of electron neutrino survival probability,
averaged over the response of detectors, were compared with measurements.
While theoretical uncertainties in the solar model and detector response
were included in the computation\cite{Fogli} of $\chi^2$, the measurement
results given here include only experimental statistical and systematic
errors added in quadrature.  The Chlorine survival probability, from
Homestake,\cite{Homestake} is $0.332\pm 0.030$.  Gallium results\cite{N2000}
for SAGE, GALLEX and GNO were combined to give a survival probability of
$0.579\pm 0.039$.  The 5.0--20 MeV, 1258-day Super-K experimental survival
probability\cite{ref:4} is $0.451\pm 0.016$.  The best fits were with
$R\approx 58\mu m$, $mR$ around 0.0094, and a mass splitting term,
$\delta_{ee}\sim 0.84\times 10^{-7}$ eV, corresponding to
$\delta m^2\sim 0.53\times 10^{-11}$ eV$^2$ for VO.  These parameters give
average survival probabilities for Chlorine, Gallium, and water of 0.383,
0.533, and 0.450, respectively, and the energy dependence shown in
Fig.~\ref{fig:edep}.  Here the coupling between $\nu_e$
and the first KK excitation replaces $\sin^22\theta$ by $4m^2R^2 = 0.00035$.

Vacuum oscillations between the lowest two mass eigenstates nearly eliminate
electron neutrinos with energies of 0.63 MeV/$(2n+1)$ for $n = 0, 1, 2, \dots$
Thus Fig.~\ref{fig:edep} shows nearly zero $\nu_e$ survival near 0.63 MeV,
partly eliminating the $^7$Be contribution at 0.862 MeV, and giving a dip at
the lowest neutrino energy.  Note that the pattern of two eigenstates very
close in mass persists for the Kaluza-Klein excitations as well.  These MSW
resonances start causing the 3rd and 4th eigenstates to be
significantly occupied above $\sim 0.8$ MeV, the 5th and 6th eigenstates
above $\sim 3.7$ MeV, the 7th and 8th above $\sim 8.6$ MeV, and the 9th
and 10th above $\sim 15.2$ MeV.
Fig.~\ref{fig:edep} shows dips in survival
probability just above these energy thresholds.

\begin{figure}[th]
\epsfxsize=3in
\begin{center}
\leavevmode
\epsfbox{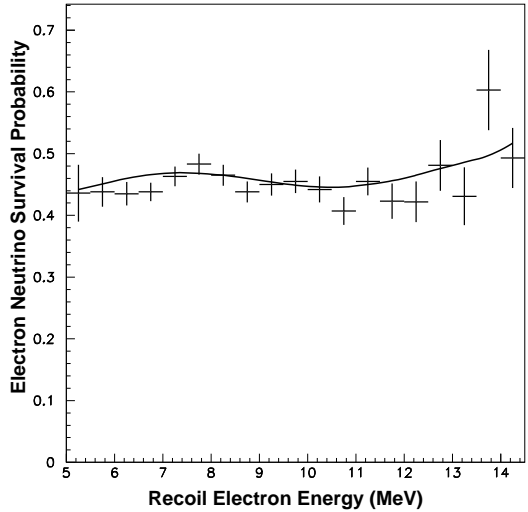}
\end{center}
\caption{Super-Kamiokande energy spectrum: measured\protect\cite{ref:4}
results based on 1285 days (error bars)
and predicted (curve) for the same parameters as in
Fig.~\protect\ref{fig:edep}.
The curve is not a fit to these data.}
\label{fig:skspect}
\end{figure}

The expected energy dependence of the $\nu_e$ survival probability is
compared with Super-K data\cite{ref:4} in Fig.~\ref{fig:skspect}.  The
uncertainties are statistical only.  The parameters used in making
Fig.~\ref{fig:skspect} were chosen to provide a good fit ($\chi^2=3.4$)
to only the total rates; they were not adjusted to fit this spectrum.
Combining spectrum data with rates\cite{GHPV} gives $\chi^2=14.0$
for the spectrum predicted from the fit to total rates.  With 18 degrees of
freedom, the probability of $\chi^2>14.0$ is 73\%.  If instead the fit were to
an undistorted energy spectrum the $\chi^2$ would be 19.0.  If VO were
eliminated, the best fit to the rates gives $\chi^2=4.4$, whereas the same
parameters applied to the spectrum yield $\chi^2=18.7$, corresponding to a
probability of 41\%.

Despite the contribution of VO, the seasonal effects are very small and will
be hard to observe, as shown in Table~\ref{Seasonal}.  On the other hand,
the SNO experiment is about to release its energy spectrum obtained from
charged-current interactions, which give far better energy resolution than
the neutrino-electron scattering observed by Super-Kamiokande, and the
characteristic shape of the spectrum above 7 MeV should be seen if this idea
is correct.

\begin{table}[h]
\caption{Predicted seasonal variations in $\nu_e$ fluxes, excluding
the $1/r^2$ variation.  The model assumed the same parameters as were
used for Fig.~\protect\ref{fig:edep}.}
\vskip 2mm
\begin{center}
\begin{tabular}{|l|c|c|c|}
$\theta-\theta_0$ & Chlorine & Gallium & Water\\ \hline
0 (January 2)    & 0.3885   &  0.5362 & 0.4602 \\
$\pm \pi/2$      & 0.3861   &  0.5328 & 0.4600 \\
$\pi$ (July 4)   & 0.3838   &  0.5278 & 0.4598 \\
\end{tabular}
\end{center}
\label{Seasonal}
\end{table}

\section{Consequences of this Fit to the Solar Data}
The parameters required to fit the average rates of the three types of solar
neutrino experiments, if confirmed by the SNO energy spectrum, would have
some obvious consequences other than demonstrating that the four-neutrino
scheme is correct and that at least one large extra dimension exists.  For
instance, the mass eigenstate which is mainly electron neutrino is
$3\times10^{-5}$ eV, which is undetectable directly or by neutrinoless
double beta decay.  The latter process measures an effective neutrino mass,
but even the contributions to that from the $\nu_\mu$ and $\nu_\tau$ must
be sufficiently small as to make that very unlikely to be observed, although
some other conjectured processes not involving neutrino mass could cause
neutrinoless double beta decay.

The effect of the 0.06 mm extra dimension size should be detectable by
gravity experiments in the not too distant future, since the present best
limit\cite{ref:24} on such effects is less than a factor of four from that
value.  This would give experimenters a definite goal for which to design.

Such a relatively large extra dimension size raises issues about cosmological
and supernova limits from the effects of high Kaluza-Klein states of both the
sterile neutrinos\cite{ref:25} and gravitons.\cite{ref:26}  While these
constraints are necessarily somewhat suspect because the two regimes, the
hot early universe and the supernova core, are very complex and are not yet
fully understood, nevertheless if taken seriously, especially the graviton
limits may pose a problem for the low-scale model, although there are
extenuating circumstances.  Usually these constraining arguments assume a
model of $n$ dimensions, each of size $R$, which is not true of either of
our models.  For sterile neutrino limits, the phenomenology presented here
is aided because there is a single Kaluza-Klein tower based on a very small
mass, the VO $\Delta m^2$ is an order of magnitude smaller than usual, and
for MSW the equivalent $\sin^22\theta$ value is more than an order of
magnitude smaller than for standard fits.  Furthermore, for both sterile
neutrinos and gravitons the universe reheat temperature could be very low,
since anything above 0.7 MeV has cosmological validity, reducing production
of high KK states.  These are very complicated issues and under much
discussion, but should it turn out that the low-scale model does not seem
to satisfy constraints, the same phenomenology is obtained by the
theoretically more desirable high-scale model, and that appears to avoid all
of these limits, as it certainly also does if the graviton-$\nu_s$
interactions in the bulk\cite{ref:27} are a problem.

The huge density of KK states which can be produced if enough energy is
available provides an explanation of ultra high energy cosmic rays beyond
the GZK cut-off.\cite{ref:28}  Neutrinos have long been suggested\cite{ref:29}
as the source of these air showers, but providing a sufficiently large
interaction cross section has been the problem.  Achieving this without some
observable low-energy effect has been the difficulty, but these narrowly
($\sim10^{-3}$ eV) spaced KK states provide such a high density at $>10^{19}$
eV that hadronic-type cross sections can be obtained.

The means of rescuing the $r$ process described in Section~3
still works for the bulk $\nu_s$, actually assuring the adiabaticity
of the $\nu_{\mu,\tau}\to\nu_s$ level crossing.  In contrast to the
usual concern that the $\nu_s$ would provide too much supernova energy loss,
it may actually aid the blow-up of the supernova, since at early times
there is a region behind the stalled shock where the interaction potential
goes to zero, and the many KK $\nu_s$ states can reconvert to active
electron neutrinos, depositing energy just where and when it is needed.
The details of this process are being worked out with George Fuller and his
students.

Finally there is the intriguing possibility that the KK states of the sterile
neutrino may provide the main component of the dark matter.  This is also
being worked on with George Fuller and his students and is bound up with the
question of the reheat temperature; can it be high enough to produce
sufficient neutrino states without overproducing gravitons?  This appears to
be true for the high-scale model.  Some preliminary
calculations have given an interesting mix: very little hot dark matter, and
about half warm and half cold dark matter.  That combination should produce
good agreement with structure measurements over a considerable range of scale.

\section{Conclusions}
The recent reanalysis of the LSND experiment greatly strengthens the case
for three different neutrino mass differences, forcing the need for a
sterile neutrino.  It is quite remarkable that the profound problems of
producing the
heaviest elements by supernovae can be solved in a manner which requires
no adjustment of parameters if the arrangement of masses and mixings of
neutrinos is exactly that required to explain the solar $\nu_e$ deficit,
the atmospheric neutrino anomaly, and the observations of the LSND experiment
(or alternatively the need for hot dark matter).  On the other hand, this
apparently successful
four-neutrino scheme fails quantitatively (as do other models) to explain
all the solar $\nu_e$ data, unless the essential sterile neutrino is a bulk
neutrino of extra large dimensions.  The resulting Kaluza-Klein tower of
states provides both MSW and vacuum oscillations, explaining the otherwise
confusing solar data.  This excellent fit to the data may be providing the
first experimental evidence for large extra dimensions, but the SNO energy
spectrum should settle this issue soon.  A positive result will have
wide-ranging consequences for gravity experiments, double beta decay,
ultra-high-energy cosmic rays, supernovae, and dark matter.

\section{Acknowledgments}
The author thanks R.N.~Mohapatra for joint early work on the neutrino scheme,
as well as the recent collaboration on the issue of extra dimensions, for
which S.J.~Yellin is also thanked.  I am grateful also to G.M.~Fuller and
Y.-Z.~Qian, with whom the heavy-element nucleosynthesis work was done, and to
Fuller, K.~Abazajian, and M.~Patel for on-going supernova and dark matter
collaborative effort.  My
work was partially supported by the U.S.~Department of Energy under contract
DE-FG03-91ER40618.

\def\Journal#1#2#3#4{{\it #1} {\bf #2} (#4) #3}

\def\APJ{Astrophys. J.}
\def\APJS{Astrophys. J. Suppl.}
\def\APP{Astropart. Phys.}
\def\ARAA{Ann. Rev. Astron. Astrophys.}
\def\ASJ{Astron. J.}
\def\IBID{ibid.}
\def\NAT{Nature}
\def\NCA{Nuovo Cimento}
\def\NIM{Nucl. Instrum. Methods}
\def\NIMA{Nucl. Instrum. Methods A}
\def\NIMPRA{Nucl. Instr. and Meth. in Phys. Res. A}
\def\NJP{New Jour. Phys.}
\def\NPA{Nucl. Phys. A}
\def\NPB{Nucl. Phys. B}
\def\NPBPS{Nucl. Phys. B (Proc. Suppl.)}
\def\PLB{Phys. Lett.  B}
\def\PPNP{Prog. Part. Nucl. Phys.}
\def\PRC{Phys. Rev. C}
\def\PRD{Phys. Rev. D}
\def\PREP{Phys. Rep.}
\def\PRL{Phys. Rev. Lett.}
\def\RMP{Rev. Mod. Phys.}
\def\SCI{Science}
\def\SJNP{Sov. J. Nucl. Phys.}
\def\ZPC{Z. Phys. C}

\def\pl{Phys. Lett.}
\def\prd{Phys. Rev. D}

\end{document}